\documentclass{article}

\usepackage{arxiv}
\usepackage[utf8]{inputenc} 
\usepackage[T1]{fontenc}    
\usepackage[hidelinks,colorlinks=true,linkcolor=blue,citecolor=blue]{hyperref}
\usepackage{url}            
\usepackage{booktabs}       
\usepackage{amsfonts}       
\usepackage{nicefrac}       
\usepackage{microtype}      
\usepackage{lipsum}		
\usepackage{graphicx}
\usepackage{natbib}
\usepackage{doi}
\usepackage{multirow}
\usepackage{float}
\usepackage{amsmath}

\title{
The maximum mass and deformation of rotating  strange quark stars with strong magnetic fields}

\author{\href{https://orcid.org/0000-0003-2835-3652}{\includegraphics[scale=0.06]{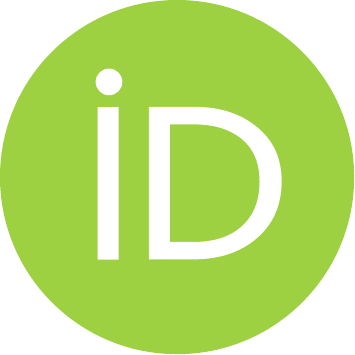}\hspace{1mm}Fatemeh  Kayanikhoo} \\
	Nicolaus Copernicus Astronomical Center \\Polish Academy of Sciences, \\ Bartycka 18, 00-716 Warsaw, Poland \\
	\texttt{fatima@camk.edu.pl} \\
	\And
     Mateusz Kapusta \\
	Astronomical Observatory, University of Warsaw, \\ Al. Ujazdowskie 4, 00-478 Warszawa, Poland\\
        \And
	\href{https://orcid.org/0000-0002-3434-3621}{\includegraphics[scale=0.06]{orcid.pdf}\hspace{1mm}Miljenko \v{C}emelji\'{c}} \thanks{Nicolaus Copernicus Astronomical Center of the Polish Academy of Sciences, Bartycka 18, 00-716
Warsaw, Poland and Academia Sinica, Institute of Astronomy and Astrophysics, P.O. Box 23-141,
Taipei 106, Taiwan} \\
	Research Centre for Computational Physics and Data Processing,\\ Institute of Physics, Silesian University in Opava,\\ Bezru\v{c}ovo n\'am.~13, CZ-746\,01 Opava, Czech Republic \\
 }
\date{\today}

\begin{document}
\maketitle

\begin{abstract}
We study the structure and total energy of a strange quark star (SQS) endowed with a strong magnetic field with different rotational frequencies. The MIT bag model is used, with the density-dependent bag constant for the equation of state (EOS). The EOS is computed considering the Landau quantization effect regarding the strong magnetic fields (up to $5\times10^{17}$ G) in the interior of the strange quark star. Using the LORENE library, we calculate the structural parameters of SQS for different setups of magnetic field strengths and rotational frequencies. In each setup, we perform calculations for $51$ stellar configurations, with specified central enthalpy values. We investigate the configurations with the maximum gravitational mass of SQS in each setup. Our models of SQSs are compared in the maximum gravitational mass, binding energy, compactness, and deformation of the star. We show that the gravitational mass might exceed $2.3 M_\odot$ in some models, which is comparable with the mass of the recently detected ``black widow'' pulsar \emph{PSR J0952-0607} and the mass of \emph{GW190814} detected by the LIGO/Virgo collaboration. The deformation and maximum gravitational mass of SQS can be characterized by simple functions that have been fitted to account for variations in both magnetic field strength and frequency. Rapidly rotating strange stars have a minimum gravitational mass given by the equatorial mass-shedding limit.
\end{abstract}

\keywords{Strange quark star, Equation of state, Magnetic field, MIT bag model, LORENE library}

\maketitle

\section{\label{sec:level1}Introduction}
 At extremely high densities ($\geq 10^{15}$ $\mathrm{g/cm^3}$), which might be found in the cores of compact objects like neutron stars, the strong force that holds quarks together within hadrons can weaken to the point that the quarks are no longer confined within these particles. Instead, they form a dense quark matter phase in which strange quarks can exist. The stability of strange quark matter (SQM) regarding the comparable energy per baryon (E/A) with the value of $E/A$ ($^{56}Fe$) $\cong 930$ $\mathrm{MeV}$ confirms that it might be the stable type of matter \cite{Bodmer_1971, Terazawa_1977, Witten_1984, farhi1984strange}. 
 
 Among many suggested kinds of objects containing SQM, a strange quark star (SQS) and a hybrid star are often considered. Strange quark stars are hypothetical compact objects that are made entirely of SQM, while hybrid stars are composed of a quark matter core surrounded by a shell of hadronic matter. The study of these exotic types of compact objects is of great interest, as they can provide insights into the properties of matter at extreme densities and temperatures. The possibility of the existence of SQSs for the first time was studied independently in \citet{Alcock_1986} and \citet{Haensel_1986}. They studied the stability of SQSs and the stellar parameters of these objects compared to the neutron stars. \citet{Weber_1997} discussed a quark deconfinement in the core of neutron stars. According to their model, the nuclear matter in the outer layers of the star is composed of protons and neutrons, while the quark matter in the central core is made up of quarks and gluons.
 
Scenarios for the phase transition of nuclear matter to the quark matter in the core of compact objects depend on the astrophysical situation. 
One possibility is a binary system containing a neutron star (NS) or a proto-NS, wherein the companion star is overflowing its Roche lobe and experiencing accretion \cite{Ouyed_2002, Ouyed_2013, Ouyed_2015}. 
Another possibility is a decrease in the centrifugal force when an NS spins down. In both those scenarios the associated gravitational implosion may be accompanied by a luminous ejection of the envelope that is called a Quark-Nova and the remnant core might be an SQS \cite{Ouyed_2002}. Conversion of an NS to an SQS releases the energy in order of $10^{52}$ ergs \cite{ouyed2022}. The possibility that quark-novae explosions are possible sources for the emission of X-rays, gamma-rays, and fast radio bursts (FRBs) observed in the universe is explored in \cite{Ouyed_2020_GRB}. 
Another observational signature, proposed to indicate the presence of a strange star, is the recent discovery of an object in the supernova remnant \emph{HESS J1731-347}. Through analysis of the X-ray spectrum and use of distance estimates obtained from Gaia observations, it is possible to estimate the object's mass and radius as $M_g= 0.77 ^{+0.20}_{-0.17} M_{\odot}$ and $R = 10.40 ^{+0.86}_{-0.78}$ $\mathrm{km}$, respectively \cite{Doroshenko_2022}. These findings suggest that the observed object may either be the lightest known neutron star, or an SQS characterized by an exotic equation of state.

In order to provide a more realistic model for compact objects one should consider the properties of these objects such as composition, magnetic field strength, and rotation of the object. The equation of state (EOS) of compact objects is an open problem in astrophysics. There are an enormous number of EOS models for compact objects. Some well-known models are MIT bag model \cite{Johnson:1975zp,Hecht2000} and NJL model \cite{Prov_2009_NJL,Prov_2020_Hyb,Prov_2020_hyb_njl} and CDDM model for SQS \cite{Chakrabarty:1989bq,CHAKRABARTY1989112,Hou_2015}. The choice of the EOS model can have a significant impact on the predicted properties of compact objects, such as mass, radius, and moment of inertia. Comparing these predictions to observational data, such as pulsar timing and gravitational wave observations, can help to constrain the EOS and provide insight into the nature of matter under extreme conditions. LIGO-VIRGO collaborations detected gravitational waves from the compact objects which carry important information about the interior material and shape of compact objects \cite{Dem:2010:Nature:,Zhao:2015:,Abbott:2020:}. 

\begin{figure}
 \centering
  \includegraphics[width=0.7\textwidth]{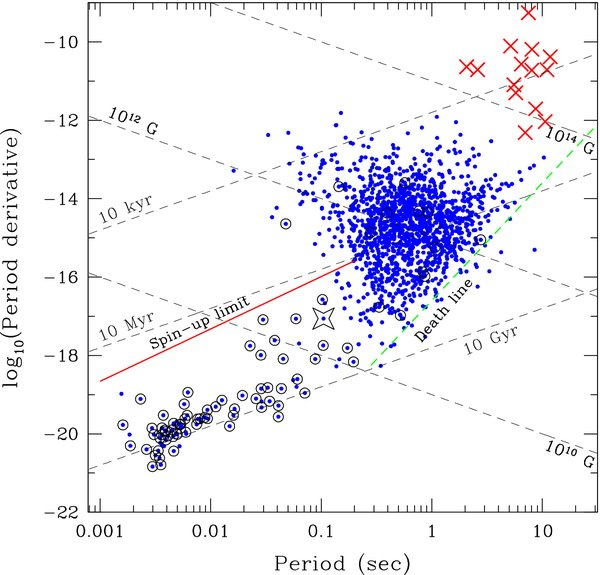}
 \caption{The P$\mathrm{\dot{P}}$-diagram of isolated pulsars (dots), binary radio pulsars (circled dots), and magnetars (crosses). Figure adapted from \cite{Halpern_2009}.}
 \label{PP}
\end{figure}

In the P$\mathrm{\dot{P}}$-diagram in Fig. \ref{PP}, shown are the magnetic field, period, and the age of objects \cite{Halpern_2009}. The distribution of magnetars is shown with the red crosses in the right top corner of the panel, isolated pulsars are shown with dots and binary pulsars are shown with circled dots. The stronger magnetic field is more spinning down the fast-spinning nascent magnetars so that at present they are rotating slowly.
In Fig. \ref{PP}, it is shown that the surface magnetic field of magnetars is $\geq 10^{14}$ G with a spin period of $\sim 10$ s, and isolated pulsars have a surface magnetic field of $\sim 10^{10}-10^{14}$ G with a spin period of $\sim 0.1-10$ s. The lowest magnetic fields correspond to the millisecond pulsars in binary systems with a spin period of $10^{-3}-10^{-1}$ s. 

According to the theoretical studies the magnetic field in the core of pulsars and magnetars might reach $10^{18}$ G \cite{Haensel_1986,Lai_1991,Boc_1995,Isa_2014}. There are a few studies on the stellar properties of magnetized neutron stars and strange stars \cite{Mallick_2014,Chatterjee_2015,Mastrano_2015}. They studied the stellar parameters such as maximum gravitational mass, radius, and deformation of NS affected by the strong magnetic field. The other considerable property of compact objects which affects the dynamic and configuration of the star is spin. Rotating compact objects are supposed to be stable with larger maximum gravitational masses compared to the non-rotating models. The rapidly rotating SQS is studied by \citet{Gondek-Rosinska_2000}. They showed that compared to models of neutron stars, the effect of rotation has a more significant impact on the overall parameters of strange stars. The Keplerian frequency of strange stars is studied by \citet{Haensel_2009}. 

In this paper, we study the effect of stellar spin on the parameters of magnetized SQS. We are interested in estimating the highest possible rotational frequency for the magnetized SQS and the effect of stellar spin and magnetic field strength on the shape and dynamical properties of SQS. We investigate different stellar models: non-rotating non-magnetized SQS, non-rotating magnetized SQS, rotating non-magnetized SQS, and rotating magnetized SQS. The strength of the magnetic field is up to $5\times 10^{17}$ G and the chosen rotational frequencies are $0$, $400$, $800$, and $1200$ Hz. 

In the second section of this paper, we present the EOS model of SQS, and in \S\ref{STR} we derive the structure equation of the star in axisymmetric space-time. We introduce the numerical code LORENE and give the numerical setup for the SQS model in \S\ref{NR}. The stellar parameters such as gravitational mass, radius, and stability as well as the total energy, binding energy, and compactness of SQS are investigated in Section \S\ref{RES}. 

\section{Equation of state}\label{sEOS}
We consider an SQS which contains up, down and strange quarks. The mass of the strange quark is $150$ MeV. The fraction of electrons is low, $10^{-3}$, so we simplify our model by neglecting their contribution. The EOS is computed in the MIT bag model \cite{Johnson:1975zp, Hecht2000} with density-dependent bag constant $\mathcal{B}_{\mathrm{bag}}$. In this model, the total energy contains the kinetic energy of quarks that is computed from Fermi relations and the bag constant:
\begin{equation}
\varepsilon_{\mathrm{tot}}=\sum_{i, j=\pm}\varepsilon_{i}^{(j)}+\mathcal{B}_{\mathrm{bag}},		\label{E03}
\end{equation}
where $j=\pm$ is the spin of quarks and $i\in (1,2,3)$ represents $up$, $down$, and $strange$ quarks.
Due to the strong magnetic field interior of the compact object, we rewrite the Fermi relations considering the Landau quantization effect \cite{Lan:1977:QM:,Lop:2015:JCAP:,Muk:2017:RRP:}. 

The single particle energy density is,  
\begin{equation}
\epsilon^{i}=[p_{i}^{2}c^{2}+m_{i}^{2}c^{4}(1+2 \nu B_{D})]^{1/2},		\label{E01}
\end{equation}
where $p_{i}$ and $m_{i}$  are the momenta and the mass of quarks, the Landau levels are denoted by $\nu$ and the dimensionless magnetic field is defined as $B_{D}=B/B_{C}$, where $B_{C}=m_{i}^{2}c^{3}/q_{i} \hbar$, with $q_{i}$ the charge of quark $i$. 

The number density of quarks is obtained as follows,
\begin{equation}
\rho=\sum_{\nu=0}^{\nu_{max}} \frac{2qB}{h^{2}c}g(\nu) p_{F}(\nu) 		\label{E02}
\end{equation}
where $\nu_{max}$ is the maximum number of Landau levels corresponding to the maximum Fermi energy $\epsilon_{\mathrm{Fmax}}$,
\begin{equation}
    \nu_{\mathrm{max}}=\frac{\epsilon_{\mathrm{Fmax}}^{2}-1}{2 m_{i} c B_{D}} \label{E02_1},
\end{equation}
in Eq. \ref{E02} $g(\nu)$ and $p_{F}(\nu)$ are the degeneracy and Fermi momentum of the $\nu$-th Landau level.
The kinetic energy density of particles is defined as
\begin{equation}
\varepsilon_{i}^{(j)}=\frac{2B_{D}}{(2\pi)^{2}\lambda^{3}}m_{i}c^{2}\sum_{\nu=0}^{\nu_{\mathrm{max}}} g_{\nu}(1+2\nu B_{D})\eta(x).
\label{E04}
\end{equation}
where
\begin{equation}
\eta(x)=\frac{1}{2}\left[x\sqrt{1+x^{2}}+ln(x\sqrt{1+x^{2}})\right].
\label{E05}
\end{equation}
with
\begin{equation}
x=\frac{X^{(j)}_{F}}{(1+2\nu B_{D})^{1/2}},		\label{E06}
\end{equation}
and 
\begin{equation}
X^{(j)}_{F}=(\epsilon_{F}^{(j)2}-1-2\nu B_{D})^{1/2}		\label{E07}
\end{equation}

Considering the zero magnetic field, $\nu_{\mathrm{max}}\to \infty$, so that the kinetic energy is simplified to the Fermi relation.

The density-dependent bag constant $\mathcal{B}_{\mathrm{bag}}$ is defined with a Gaussian relation, 
\begin{equation}
\mathcal{B}_{\mathrm{bag}}(\rho)=\mathcal{B}_{\mathrm{\infty}}+(\mathcal{B}_{0}-\mathcal{B}_{\mathrm{\infty}})e^{-\alpha(\rho/\rho_0)^2},
\label{E08}
\end{equation}
with $\alpha = 0.17 $ and $\mathcal{B}_{0}=\mathcal{B}_{bag}(0)=400~\mathrm{MeV/fm^{3}}$. We should define $\mathcal{B}_{\mathrm{\infty}}$ in such a way that the bag constant would be compatible with experimental data (CERN-SPS). We determine $\mathcal{B}_{\mathrm{\infty}} =8.99~  \mathrm{MeV/fm^{3}}$ by putting the quark energy density equal to the hadronic energy density \cite{LOCV2000}.

\begin{figure}
 \centering
  \includegraphics[width=0.8\textwidth]{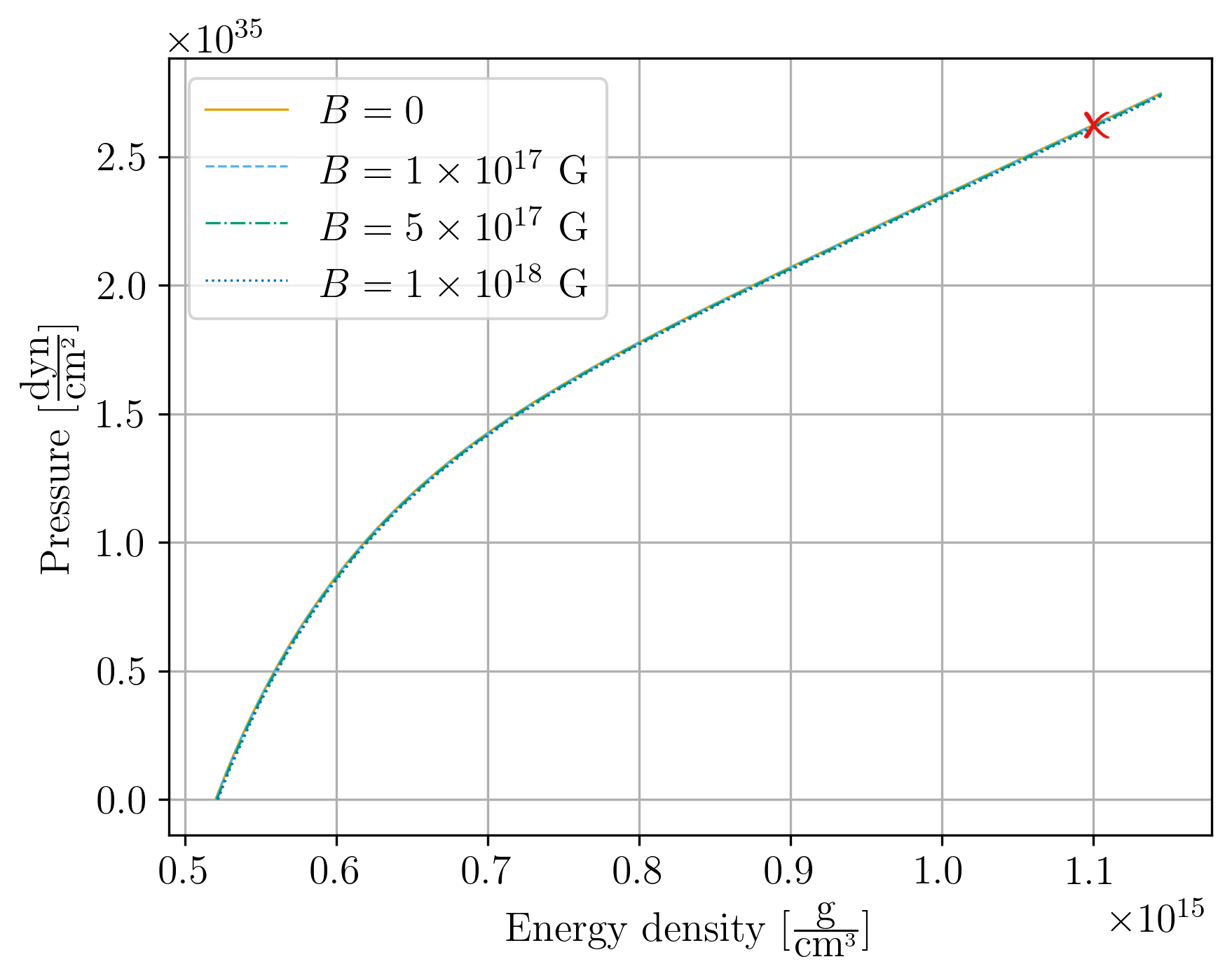}
 \caption{Pressure as a function of the energy density of SQM in the presence of magnetic fields of different strengths. The energy density corresponding to the maximum gravitational mass is pointed with the red cross. The curves with four magnetic fields overlap.}\label{EOS}
\end{figure}

The EOS is given by
\begin{equation}
P(\rho)=\rho \left(\frac{\partial \varepsilon_{tot}}{\partial \rho}\right) - \varepsilon_{tot}.		\label{E09}
\end{equation}
Pressure versus the energy density in the presence of magnetic fields of different strengths in SQM is shown in Fig.\ref{EOS}. Our model of EOS indicated zero pressure at the energy density of $\approx 0.52 \times 10^{15}$ $\mathrm{gr/cm^3}$ ($\approx 290 \mathrm{MeV/fm^3}$), and is continued to the energy density corresponding to the maximum mass of SQS, which is approximately pointed with a red cross on Fig. \ref{EOS}. Although the effect of the magnetic field strength on EOS is negligible, in the following sections we show that the structural parameters and shape of the star change significantly. 

A number of studies have investigated different models of EOS for quark stars. To assess the performance of our model, we compared it to the EOS presented by \citet{Chatterjee_2015}. They provided the linear EOSs in which the maximum mass is inversely proportional to the square root of energy density at zero pressure. Our EOS model is approximately linear at higher densities and the extrapolation of our EOS shows the energy density $\approx 2\times 10^{14}$ $\mathrm{g/cm^3}$ that is half of the value of the model provided by \citet{Chatterjee_2015}. The larger maximum gravitational mass is expected from our EOS model. They also confirm that the magnetic field less than $10^{19}$ G does not significantly impact the stiffness of EOS.  

\section{The equations of stellar structure}\label{STR}
In this section, we derive the differential equations of the stellar structure by solving Einstein's equation,
\begin{equation}
    R^{\mu \nu} - \frac{1}{2} R g^{\mu \nu} = 8 \pi T^{\mu \nu}
\end{equation}
where $R^{\mu \nu}$ is the Ricci tensor, $R$ is the Ricci scalar, $g^{\mu \nu}$ is the metric coefficient and $T^{\mu \nu}$ is the energy-momentum tensor. We choose units with $G=c=1$. 

We start with the energy-momentum tensor of the perfect fluid and the spherically symmetric star. Then, regarding the strong magnetic field in the microscopic properties of SQS, we inspect the energy-momentum tensor coupling with Maxwell energy-momentum tensor.

Tolman-Oppenheimer-Volkov
equations (TOV) are derived by solving the Einstein field equations in the spherically symmetric space-time and the perfect fluid's energy-momentum tensor \cite{Tolman_1939, Open_Vol_1939}, 
\begin{equation} \label{perfect}
    T^{\mu \nu} = (\varepsilon+P)u^\mu u^\nu + P g^{\mu \nu}
\end{equation}
where $u^{\mu}$ is the fluid 4-vector, $\varepsilon$ is the energy density and $P$ is the pressure of the perfect fluid.
We can write
\begin{equation} \label{TOV}
    \frac{dP}{dr} = -(P+\varepsilon)\frac{m+4 \pi r^3 P}{r(r-2m)}
\end{equation}
and
\begin{equation}
    \frac{dm}{dr} = 4 \pi r^2 \varepsilon .  
\end{equation}
In the presence of the magnetic field, the interaction of the electromagnetic field with the matter (magnetization) is considerable, the energy-momentum tensor is given by
\begin{equation}
    T^{\mu \nu}=(\varepsilon + P)u^{\mu}u^{\nu}+P g^{\mu \nu}+\frac {\mathcal{M}}{B}\Big[ b^{\mu}b^{\nu}-(b\cdot b)\\(u^{\mu}u^{\nu}+g^{\mu \nu}) \Big]+
\frac{1}{\mu_{0}}\Big[-b^{\mu}b^{\nu}+(b\cdot b)(u^{\mu}u^{\nu}+\frac{1}{2}g^{\mu \nu})\Big], 	\label{E12}
\end{equation}
where the two first terms are the perfect fluid contribution (Eq. \ref{perfect}), the third term is the magnetization contribution and the last term is the pure magnetic field contribution to the energy-momentum tensor. $B$ is the magnetic field, and $b^{\mu}$ is the magnetic field 4-vector. $\mathcal{M}$ represents the interaction of the electromagnetic field with the matter, which is given by the coupling between the electric current $j^\phi$ and the magnetic vector potential, 
\begin{equation}
   j^\phi=\Omega j^t + (\varepsilon+P)k_0,
\end{equation}
where $\Omega$ is rotational velocity of the star and $k_0$ is the current function.

We solve the Einstein field equations within the 3+1 formalism in a stationary, axisymmetric space-time \cite{Chatterjee_2015,Franzon_2017}. The metric is given by
\begin{equation}
    ds^{2}=-N^2 dt^2+A^2(dr^2+r^2 d \theta^2)+\lambda^2r^2\sin^2(\theta)(d\phi-N^{\phi}dt)^2		\label{E10}
\end{equation}
where $N$, $A$, $\lambda$, and $N^{\phi}$ are functions of $(r,\theta)$.
By applying 3+1 formalism we obtain a set of four elliptic partial differential equations,
\begin{equation}
\Delta_3=4\pi A^2(E^T+S^{r}_{r}+S^{\theta}_{\theta}+S^{\phi}_{\phi})+\frac{\lambda^2 r^2 \sin^2(\theta)}{2N^2}\delta N^{\phi}\delta N^{\phi}\\
-\delta \nu \delta(\nu+\beta)		\label{E11a}
\end{equation}
\begin{equation}
\Delta_2[\alpha+\nu]=8\pi A^2 S^{\phi}_{\phi}+\frac{3\lambda^2 r^2 \sin^2(\theta)}{4N^2}\delta N^{\phi}\delta N^{\phi}-\delta \nu \delta \nu 		\label{E11b}
\end{equation}
\begin{equation}
\Delta_2[(N\lambda -1)r \sin(\theta)]=8\pi N A^2 \lambda r \sin(\theta)(S^{r}_{r}+S^{\theta}_{\theta})		\label{E11c}
\end{equation}
and
\begin{equation}
\left[\Delta_3-\frac{1}{r^2 \sin^2(\theta)}\right](N^{\phi}r \sin(\theta))=-16\pi \frac{NA^2}{\lambda^2}\frac{J^{\phi}}{r \sin(\theta)}+\\
r \sin(\theta)\delta N^{\phi}\delta(\nu-3 \beta),	\label{E11d}
\end{equation}
where $\nu=\ln N$, $\alpha=\ln A$, $\beta=\ln \lambda$, and $J^{\phi}$ is electromagnetic current. In the above equations, $E^T$ and $S^i_j$ are total energy and stress, respectively.

 Considering a magnetic field pointing in the z-direction, we can rewrite the energy-momentum tensor in a well-known form:
 
\begin{equation}
T^{\mu \nu}= \\
\mathrm{diag}\left(\varepsilon +\frac{B^{2}}{2\mu_{0}}
, P-\mathcal{M}B+\frac{B^{2}}{2\mu_{0}}, P-\mathcal{M}B +\frac{B^{2}}{2\mu_{0}}, \right.\\
\left. P-\frac{B^{2}}{2\mu_{0}} \right).
\label{E14}
\end{equation}
In this equation, the magnetization term reduces the total pressure of the system. It is also clear that the magnetic field reduces the parallel pressure, but the perpendicular pressure increases with increasing the magnetic field.

\section{The numerical method}\label{NR}
We solve a set of four elliptic partial differential equations presented in Section \S\ref{STR} using LORENE library (http://www.lorene.obspm.fr) \cite{Bonazzola_1998,Chatterjee_2015,Franzon_2017}. By employing spectral methods, LORENE provides a more accurate approach to solving partial differential equations than grid-based methods, thereby enabling more precise calculations of the solutions to this system. Space in LORENE is separated into domains and mapped onto the specific coordinate system that can be readjusted in order to tackle non-spherical shapes. Our setup consists of $3$ domains: two inside and next to the surface of the star and one outside at infinity. We use Et\_magnetisation class to calculate hydrostatic configurations for uniformly (not differentially) rotating magnetized stars (the code is located in ``\emph{Lorene/Codes/Mag\_eos\_star}''). 

To specify the magnetic field, LORENE uses the so-called current function, $k_0$, which describes  the amplitude of current inside the star to generate the magnetic field. In our setup, the current function amplitude changed from $0$ to $15000$ in the intervals of $2000$, enabling us to cover vast ranges of central magnetic field values to see how SQS behaves even in the fields up to $5\times 10^{17}$ G. As we mentioned in the Introduction, we solve the equations for different rotational frequencies. In every series of calculations, we compute the parameters of $51$ stellar configurations with specified central enthalpy values in the range from $0.01$ $\textrm{c}^2$ to $0.51$ $\textrm{c}^2$ with the spacing of $0.01$ $\textrm{c}^2$. 

To expedite the calculations, we utilized a wrapper based on MPI that facilitates the distribution of the computational load across the available threads. By such parallelizing, we significantly enhanced the speed and efficiency of the calculations, resulting in faster processing times and improved performance. 

Equilibrium configurations in Newtonian gravity are known to satisfy the virial relation when a polytropic equation of state is assumed. This relation is commonly utilized to verify the accuracy of computations. The 3-dimensional virial identity (GRV3), introduced by \citet{GRV3}, extends the Newtonian virial identity to general relativity. On the other hand, the 2-dimensional virial identity (GRV2) proposed by \citet{GRV2}, generalizes the virial identity for axisymmetric space-times to general asymptotically flat space-times. Our computational results indicate a high level of accuracy which is $\approx 10^{-5}$ in the non-magnetized non-rotating models and $\approx 10^{-2}$  in the magnetized fast-rotating model.
\section{Analysis of stellar properties}\label{RES}
In this section, we examine the properties of the star under varying conditions, where the strengths of both the magnetic field and rotational frequency are modified. Through the exploration of different setups, we gain insight into the impact of these factors on the star, allowing for a more comprehensive understanding of its dynamics and properties.  

In the following four-panel figures, each panel shows the rotational frequencies ($0$, $400$, $800$, and $1200$ Hz). The color bar in all plots indicates the strength of the central magnetic field which varies from $0$ to $5\times 10^{17}$ G.

\subsection{Gravitational mass and radius}\label{MgR}
\begin{figure*}
\includegraphics[width=1\textwidth]{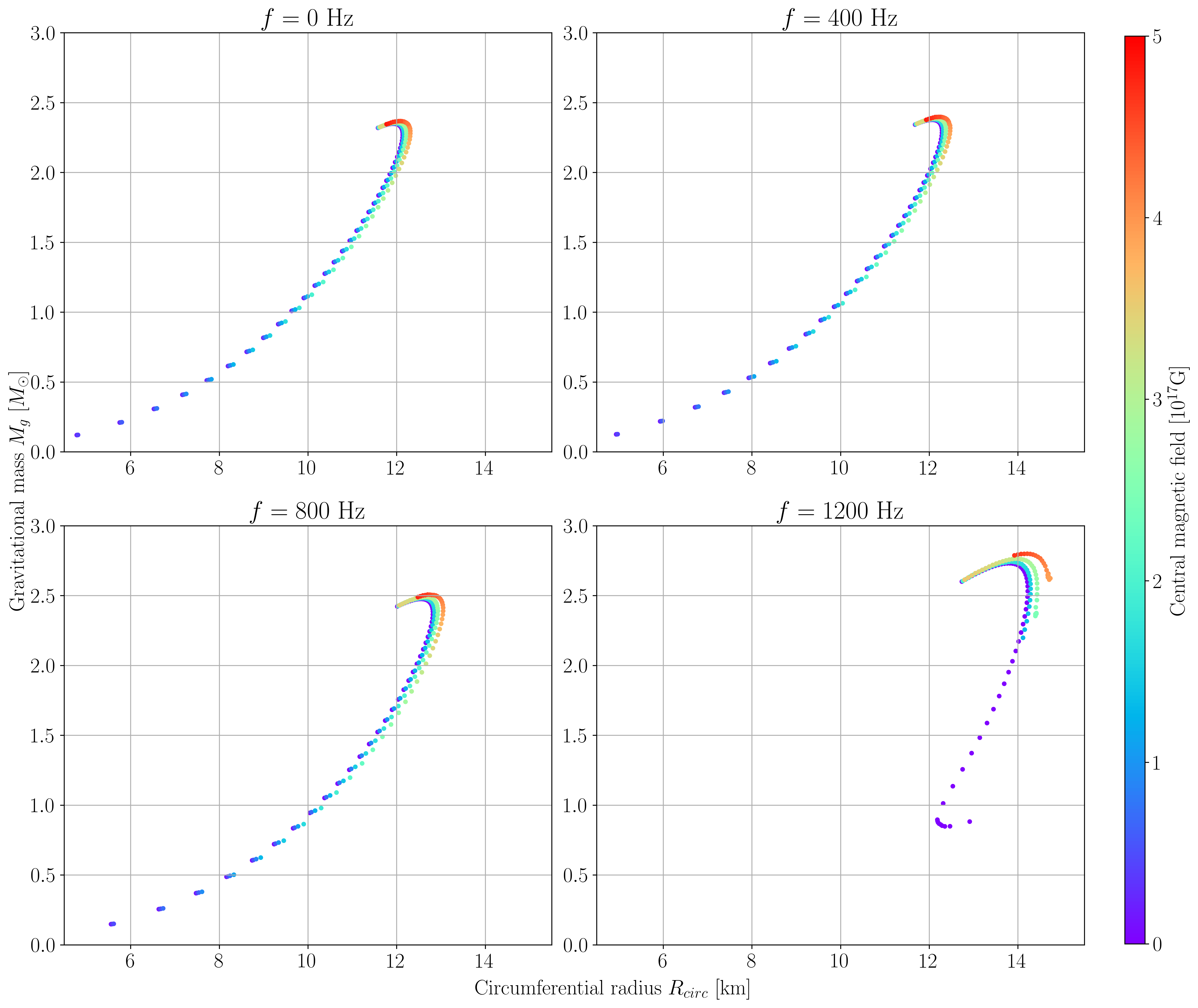}
\caption{Gravitational mass versus circumferential radius for different rotational frequencies. The color indicates the value of the central magnetic field.}\label{MR}
\end{figure*}
\begin{figure*}
\includegraphics[width=1\textwidth]{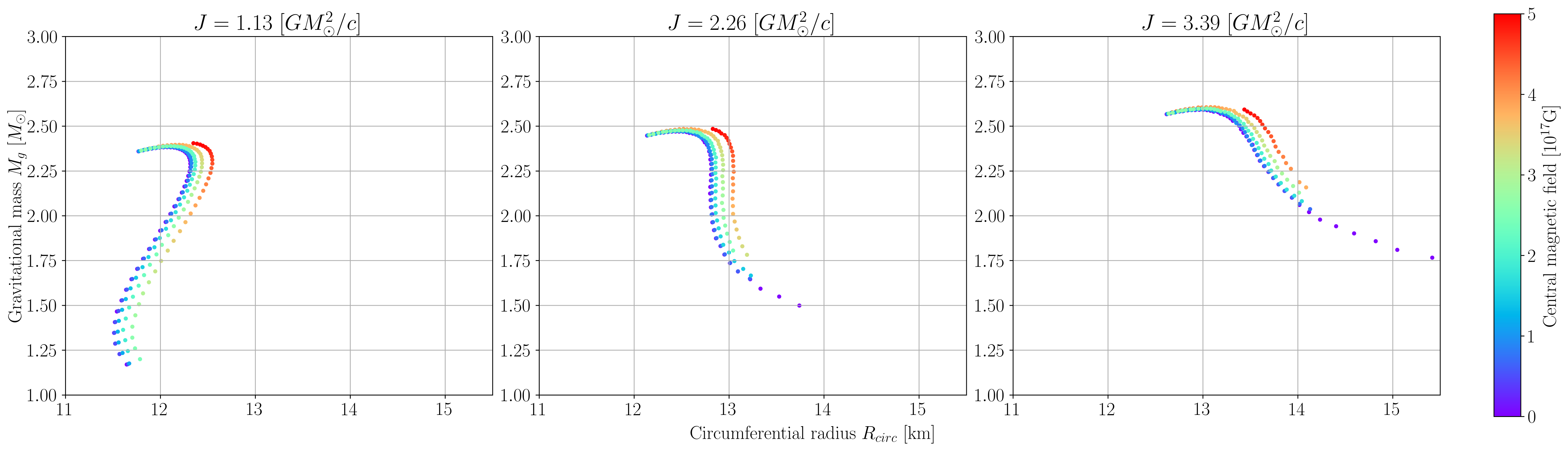}
\caption{Gravitational mass versus circumferential radius with different angular momentums. The color indicates the value of the central magnetic field.}\label{MRJ}
\end{figure*}
\begin{figure*}
\includegraphics[width=1\textwidth]{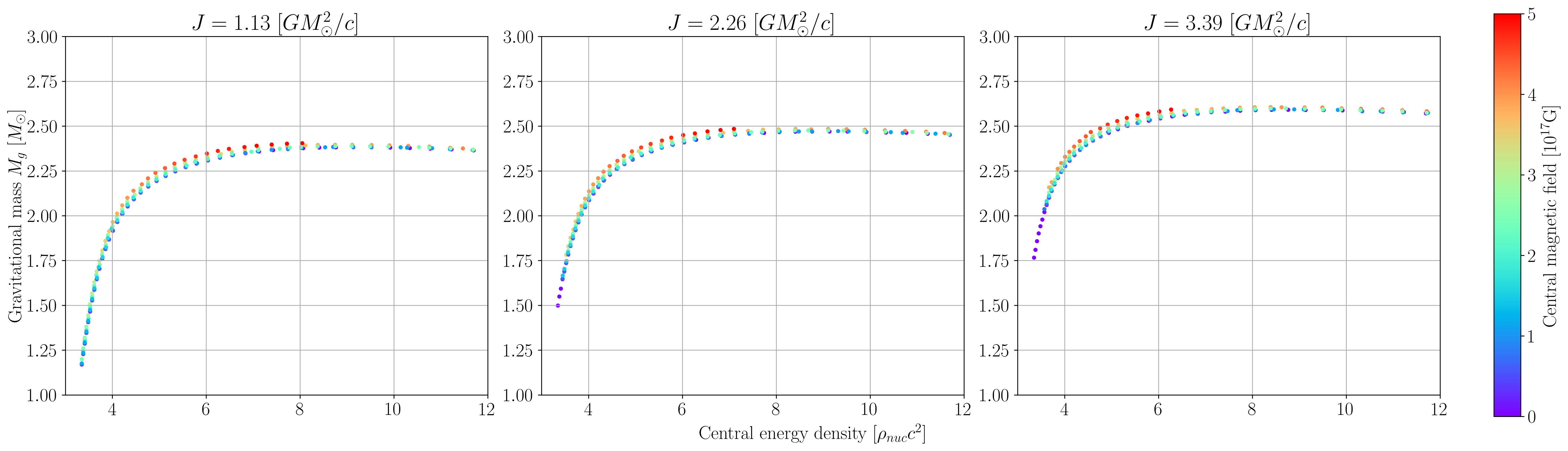}
\caption{Gravitational mass as a function of central energy density with different angular momenta. The color indicates the value of the central magnetic field.}\label{MEJ}
\end{figure*}
Mass and radius are crucial parameters for the study of compact objects, as they provide an important understanding of the underlying physics and characteristics of these objects. In Fig. \ref{MR}, we present a plot of the mass as a function of the circumferential radius, $R_{circ}$ for each configuration. For relatively small masses ($M<M_\odot$) and low rotation rates, our model obeys the mass-radius relation $M\simeq \frac{4}{3}\pi\rho_0 R^3$, characterizing self-bound stars with density $\rho_0$ at zero pressure \cite{Haensel_1986}.  
In each panel of Fig. \ref{MR}, it is shown that the maximum gravitational mass $M_g^{max}$ increases as a function of the magnetic field. $M_g^{max}$ also increases with increasing the rotational frequency. \citet{Gourgoulhon_1999} showed that the absolute maximum increase in the mass for rigidly rotating self-bound non-magnetized stars is about $44\%$, giving $M_g^{max}(\rm rot)\simeq 2.83 M_\odot$. In our model, the maximum considered frequency is $1200$ Hz, which is significantly smaller than the absolute maximum frequency for the given equation of state approximated by the formula $f_{\rm max}^{\rm EOS}=1.22 (M/M_\odot)^{1/2} R_{10}^{-3/2} {\rm Hz}$ \cite{Haensel_1995}, which gives $f_{\rm max}^{\rm EOS} \simeq 1700\,{\rm Hz}$. In our model, the increase in the maximum mass for $1200$ Hz is about $16\%$. 

The Keplerian frequency for considered stellar models can be estimated using the approximate formula $f_{\rm Kep}=1.15 (M/M_\odot)^{1/2} R_{10}^{-3/2} ~{\rm kHz}=1.2(\frac{\bar{\rho}}{5.2\,10^{14}})^{1/2}~{\rm kHz}$  and is slightly above 1200 Hz \cite{Haensel_2009}.
In the frame corresponding to $\emph{f} = 1200$ Hz, we see that only massive SQSs may exist as strongly magnetized, fast-rotating objects where the binding energy is in balance with the magnetic and rotation energy of the stars. Additionally, our  model indicates that there is a maximum limit for the magnetic field in the fast-rotating model to obtain stable configurations. 

For the stability of a rotating compact strange star, the four main constraints must be fulfilled \citep{cook1994rapidly, Gondek-Rosinska_2000}: 1) a static constraint, demanding that a solution for rotating compact (NS/SQS) object should converge to the solution for a non-rotating in the limit of zero rotation, 2) a low mass constraint, defining that an NS can not form below a certain mass limit, 3) the Keplerian constraint, by which the maximum rotation rate of a compact object can not exceed the Keplerian frequency, and 4) a stability constraint to quasi-radial perturbations, stating that a  rotating compact object should remain stable under perturbations like small changes in its shape or density distribution.

Our model meets these requirements. We provide a function defining the rotating model which, with zero rotation, converges to the non-rotating model. We approximate the equatorial mass-shedding limit. In the previous paragraphs, we discussed that the Keplerian frequency is slightly higher than the highest considered frequency in our study.  

In order to meet the fourth constraint of stability for rotating compact stars, it is necessary to investigate the stability of the model against axisymmetric perturbations. This can be done by examining the derivative of the mass with respect to the radius at constant angular momentum $J$, which is $\left( \frac{dM}{dR}\right)_J >0$. Specifically, an increase in the stellar radius at a fixed angular momentum should give increased stellar mass. This criterion indicates that the star has the ability to withstand minor deformations and oscillations without undergoing collapse or mass loss. 

To find the last stable configuration in each model, we examine the mass-radius plot at constant angular momentum $J$. An example is illustrated in Figure \ref{MRJ}, which displays three frames representing the angular momentum with $1.13$, $2.26$, and $3.30$ $\mathrm{GM_{\odot}^2/c}$. As we discussed, the maximum gravitational mass in each sequence corresponds to the last stable configuration in each magnetic field and angular momentum. We show the numerical results of the last stable configurations of SQS of each model in Table \ref{table:1}.

\begin{figure}
\centering
\includegraphics[width=0.8\textwidth]{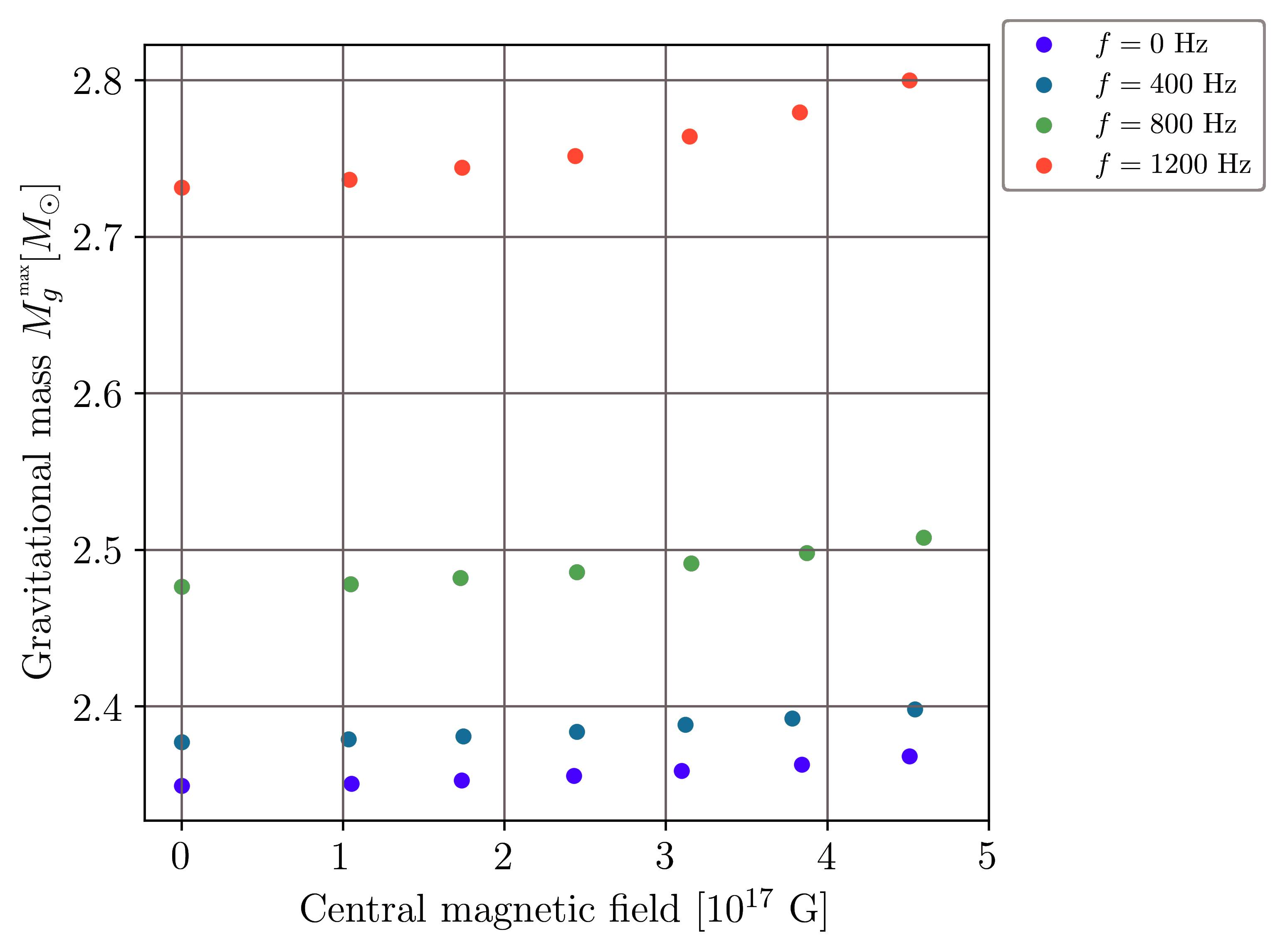}
\caption{Gravitational mass as a function of the central magnetic field at different rotational frequencies.}\label{mBf}
\end{figure}

The maximum gravitational masses ($M_g^{max}$) of models versus the central magnetic field ($B_c$) are shown in Fig. \ref{mBf}. The $M_g^{max}$ is a function of the central magnetic field $B_c$ and rotational frequency $\emph{f}$,
\begin{equation}
    \frac{M_g^{max} (B_c, f)}{M_g^{max}(0, 0)} \simeq (1+b B_c^2)(1+c f^2) \label{eqmg}
\end{equation}
where b and c are the constant values. In order to improve the accuracy of our fitting, we created a sequence of rotational frequencies ranging from $0$ to $1200$ Hz, with a step size of $200$ Hz. In result, we obtained coefficients of $c = 8.57 \times 10^{-8}$ $\textrm{s}^2$ and $b = 4.31\times10^{-38}$ ${\textrm{G}^{-2}}$. Note that the accuracy in the fitting function is greater than $10^{-3}$ in the rotational frequencies less than $1200$ Hz. 
\begin{figure}
    \centering
\includegraphics[width=0.8\textwidth]{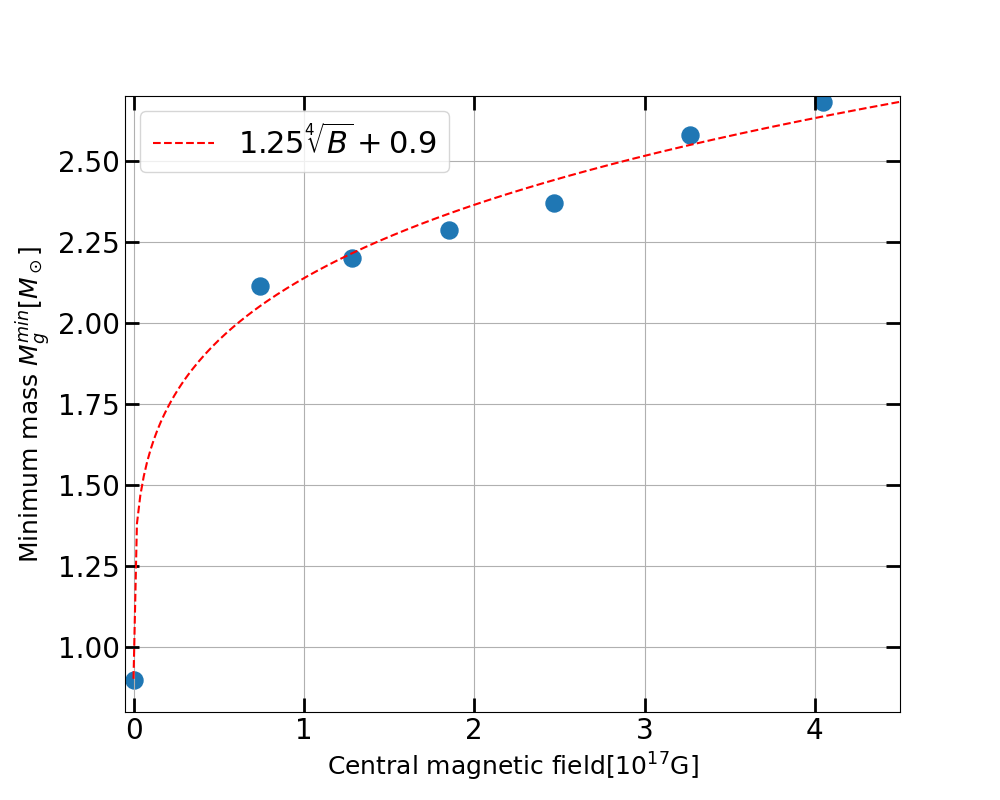}
    \caption{The minimum mass limit for $\emph{f}=1200$ Hz as a function of the central magnetic field.}
    \label{mass_shedding}
\end{figure}

We find a variation in the maximum gravitational mass ($M_g^{max}$) of non-magnetized strange quark stars (SQS) from $2.35 M_\odot$ to $2.73 M_\odot$, as the rotational frequency, rises from zero to $1200$ Hz. In the magnetized non-rotating model $M_g^{max}$ reaches $2.37 M_\odot$. However, when rotation and magnetic field are present, $M_g^{max}$ reaches $2.4 M_\odot$, $2.50 M_\odot$, and $2.80 M_\odot$ at rotational frequencies of $\emph{f}=400$, $800$, and $1200$ Hz, respectively.
It is also obvious from Fig. \ref{MgR} that there is a lower limit for the gravitational mass in each sequence of the fast-rotating stars ($f=1200$ Hz), the so-called equatorial mass-shedding limit that is discussed by \citet{Gondek-Rosinska_2000} for non-magnetized rapidly rotating quark stars. We find that the magnetic field affects the equatorial mass-shedding limit. In the rapidly rotating model with $\emph{f}=1200$ Hz, the equatorial mass-shedding changes proportional to the central magnetic field ($M_g^{min}\propto \sqrt[4]{B_c}$), as shown in Fig.~\ref{mass_shedding}. We note that accuracy of the minimum mass value presented in this figure cannot be guaranteed due to the computational challenges encountered near the critical points. The numerical error involved in these calculations may have an impact on the fitting function. 

The most recent detection of millisecond ``black widow'' pulsar \emph{PSR J09520607} estimates the gravitational mass of $M_g \simeq 2.35 M_\odot \pm 0.17$ and the dipole surface magnetic field of $B\simeq 6 \times 10^7$ G, with the period of $1.4$ s and rotational frequency of $\simeq 700$ Hz \cite{Romani_2022}. We compute the maximum gravitational mass of $M_g^{max}=2.35 M_\odot$ for the non-magnetized non-rotating SQS and $M_g^{max}=2.38 M_\odot$ for SQS with $B_c \simeq 10^{17}$ G and the rotational frequency of $400$ Hz. 
Also, our model for the pulsar with the rotational frequencies of $800$ and $1200$ Hz ($M_g^{max}\geq 2.5 M_\odot$) can explain the recent detection of \emph{GW190814} by the LIGO/Virgo collaboration, where the gravitational mass of the lower mass object in the binary is estimated between $2.5 M_\odot$ and $2.67 M_\odot$. Compact objects with masses around $\approx 2M_\odot$ have been observed also in \emph{PSR J1614-2230} ($M=1.908 \pm 0.016M_\odot$) and \emph{PSR J0348+0432} ($M= 2.01\pm 0.04 M_\odot$) \cite{Dem:2010:Nature:,Zhao:2015:}, and Chandra X-ray detection of \emph{SGR J1745–2900}, estimates the gravitational mass and radius of this magnetar up to $2M_\odot$ and $13.7$ km, the surface magnetic field of this source is $2\times 10^{14}$ G \cite{magnetar2015,magnetar2020}.

The gravitational mass as a function of central energy density for different values of magnetic field and angular momentum is shown in Fig. \ref{MEJ}. For a given central energy density, the gravitational mass increases with increasing magnetic field and angular momentum. This behavior can be attributed to the increased pressure and density gradient near the center of the star, which can support a larger mass of material. In particular, as the magnetic field strength increases, the central energy density of the maximum gravitational mass also increases. Similarly, as the angular momentum increases, there is a shift towards higher gravitational masses at a given central energy density. 
\begin{table*}
\centering
\large
\caption{Structural parameters of the configuration with the maximum gravitational mass in different models.}\label{table:1}
 \begin{tabular}{ |c ||c|c|c|c|c|c|}
  \hline
  $\emph{f}$ (Hz) & $B_c$ ($10^{17}$ G) & $M_g (M_\odot)$ & $M_b (M_\odot)$ & $R_{circ}$ (km) & $a$ &$|E_{EB}|/A  $ (MeV)\\
  \hline\hline
   \multirow{8}{8pt}{\centering 0}&0&2.35&2.92&11.92&1&184\\ 
   
   &1.05&2.35&2.92&11.9&1.0&184\\ 
       
   &2.43&2.36&2.93&11.97&1.01&184\\
    
   &3.10&2.36&2.94&12.03&1.02&184\\
    
   &3.84&2.36&2.94&12.03&1.03&184\\
    
   &4.51&2.37&2.95&12.09&1.04&183\\
    
   &5.11&2.37&2.95&12.21&1.05&183\\
   \hline\hline
  \multirow{8}{8pt}{\centering 400}&0&2.38&2.96&12.05&1.03&184\\
   
   &1.03&2.38&2.95&12.1&1.03&184\\
    
   &1.74&2.38&2.96&12.1&1.03&184\\
    
   &3.12&2.39&2.97&12.15&1.05&183\\
    
   &3.78&2.39&2.98&12.22&1.06&183\\
    
   &4.5&2.40&2.98&12.22&1.07&183\\
    
   &5.15&2.40&2.99&12.34&1.08&182\\
   \hline\hline
  \multirow{5}{8pt}{\centering 800}&0&2.48&3.07&12.54&1.11&182\\
   
   &1.05&2.48&3.10&12.55&1.12&182\\
    
   &2.45&2.49&3.10&12.59&1.13&182\\
    
   &3.87&2.5&3.10&12.65&1.15&182\\
    
   &4.6&2.51&3.11&12.70&1.16&182\\

   &5.11&2.51&3.11&12.93&1.20&180\\
   \hline \hline
    \multirow{5}{13pt}{\centering1200}&0&2.73&3.38&13.08&1.36&179\\
   &1.04&2.73&3.38&13.84&1.36&179\\
   &2.44&2.75&3.40&13.90&1.38&179\\
   &3.83&2.78&3.43&14.10&1.43&178\\
   &4.51&2.80&3.46&14.21&1.46&177\\
   &4.97&2.80&3.43&14.6&1.55&171\\
    \hline
\end{tabular}

\end{table*}
\subsection{Magnetic and rotational deformation}\label{Sdeformation}

\begin{figure*}
\includegraphics[width=1\textwidth]{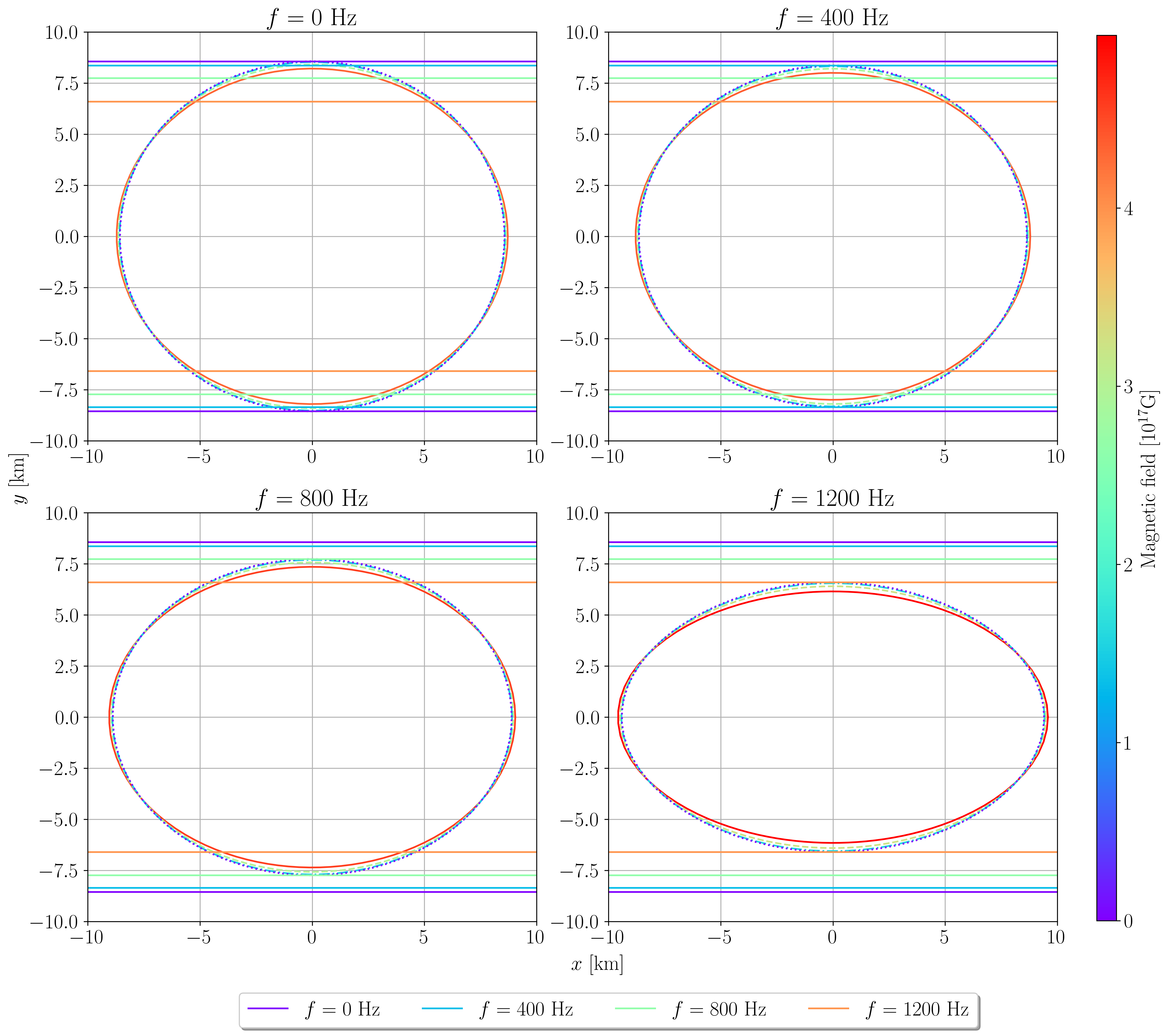}
\caption{Deformation of SQS for different rotational frequencies. The color of the ellipses indicates the magnitudes of the central magnetic field. Straight lines, which are at the same positions in all panels, indicate the polar radius of the non-magnetized configuration in each rotational frequency. 
}\label{def}
\end{figure*}
\begin{figure}
 \centering
  \includegraphics[width=0.8\columnwidth]{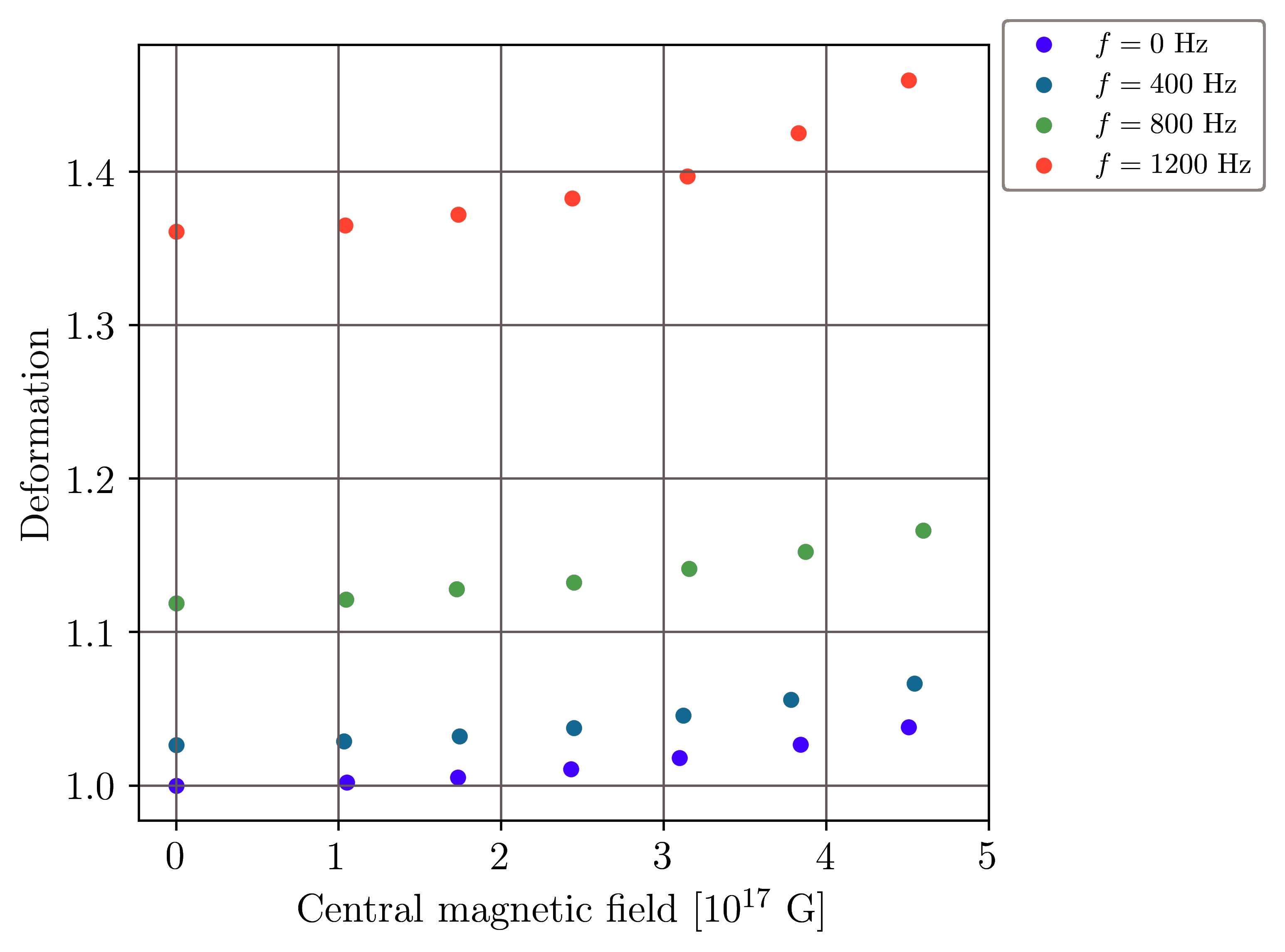}
 \caption{Deformation parameter versus the magnetic field in different rotational frequencies for the maximum stable configurations of SQS.}
 \label{Deform}
\end{figure}
Rotation and magnetic field break the spherical symmetry of stars. The magnetic deformation depends on the magnetic field configuration of the star. We consider the poloidal magnetic field, where $B_r$ and $B_\theta$ are the non-vanishing components of the magnetic field.
The magnetic and rotational deformations of SQS are shown in Fig. \ref{def}. In this figure, horizontal lines, which are at the same positions in all panels, indicate the polar radius of the non-magnetized configuration in each rotational frequency and show that rigidly rotating stars become more oblate. In each panel, colors indicate the magnetic field and we see that the strength of the magnetic field affects the shape of the star. 

We plotted the deformation parameter $a=R_{eq}/R_{pol}$ (where $R_{eq}$ is the equatorial radius and $R_{pol}$ is the polar radius) of the configurations with the maximum gravitational mass as a function of the central magnetic field in different rotational frequencies in Fig. \ref{Deform}. This helps to clarify how the magnetic field and rotational frequency affect the shape of the star. We find that the deformation parameter is a function of magnetic and rotational energy. The fitting function is
\begin{equation}
    \frac{a(B_c, f)}{a(0,0)} \simeq (1+\Tilde{b} B_c^2)(1+\Tilde{c} f^2) \label{aBf}
\end{equation}
where $\Tilde{b}=1.54\times 10^{-37}$ $\textrm{G}^{-2}$ and $\Tilde{c} = 2.39\times 10^{-7}$ $\textrm{s}^2$. Similar to Eq. \ref{eqmg}, we generated a series of rotational frequencies to improve the precision of the fitting. The accuracy of the fitting is greater than $10^{-3}$ in frequencies less than $1200$ Hz.
 
The deformation parameter $a$ can be found in the fifth column of Table \ref{table:1}. Based on our findings, we obtain that the maximum deformation parameter $a=1.55$ corresponds to magnetized spinning SQS with $B_c \simeq 5 \times 10^{17}$ G and $\emph{f} = 1200$ Hz. 
\subsection{The total energy of SQS}\label{energy}
The total energy of the star as measured at infinity is a sum of the internal and external energies: $E_{tot}=E_{int}+E_{ext}$. The model being considered consist of SQM inside the star. Outside of the star, there is a dipolar magnetic field that the strength decreases with the distance from the star.

To calculate the internal energy $E_{int}$, we integrate the energy-momentum tensor $T^{\mu \nu}$ over the volume of SQS using the LORENE code. We compute the magnetic energy outside the star $E_{ext}$, using the following method.

Given that there is only a magnetic field outside the surface of SQS, we neglect the general relativistic impacts on its external energy. The energy density of the magnetic field can be expressed as $B^2/2\mu_0$. Since $\boldsymbol{\nabla}\times{\boldsymbol{B}}=0$, we can introduce a magnetic potential $\phi$ such that $\boldsymbol{\nabla}{\phi}=\boldsymbol{B}$. This allows us to define the external energy of the star as follows: 
\begin{equation}
E_{ext}=\int_V \frac{1}{2\mu_0}\boldsymbol{\nabla} \phi\cdot\boldsymbol{\nabla} \phi =\frac{1}{2\mu_0}\int_{dV} da \left(\boldsymbol{n}\cdot\boldsymbol{\phi}\boldsymbol{\nabla} \phi \right),
\end{equation}
 where $V$ represents the domain outside of SQS. According to the Gauss theorem,
 \begin{equation}
\boldsymbol{\nabla} \cdot{\left(\phi\boldsymbol{\nabla}{\phi}\right)}=\phi\boldsymbol{\nabla}^2{\phi}+\boldsymbol{\nabla}{\phi}\cdot\boldsymbol{\nabla}{\phi}=\boldsymbol{\nabla}{\phi}\cdot\boldsymbol{\nabla}{\phi}
 \end{equation}
The triple integral can be simplified to a double integral over the surface of the star. By using the axisymmetric coordinates, this double integral can be further reduced to a single integral. As a result, we can consider $\phi$ as the only effective parameter, which can be assumed to be produced by the magnetic dipole moment $\mu$:
\begin{equation}
    \phi(r,\theta)=\frac{\mu \cos{\theta}}{4\pi r^2}.
\end{equation}
In this calculation, we used $33$ points lying on the surface of the star, sampled from LORENE, and used the Lagrange interpolation to construct the function describing the stellar surface. Then an integral over the stellar surface was calculated using the trapezoid rule.

In Table~\ref{table:2}, we show the numerical results for the external and total energy for configurations of non-magnetized, non-rotating SQS, and magnetized rotating SQSs in the highest magnetic field in each rotational frequency. The contribution of external energy to total energy is less than $1\%$. We find that the magnetic field and spin of SQS affect both gravitational mass and total energy. In result, the total energy increases $\simeq 4\%$ from non-rotating to rotating SQS. 

In our models, the total energy of the SQS is approximately $5 \times 10^{47}$ J (or $10^{54}$ ergs). For comparison, the energy of a Type II supernova is estimated to be up to $10^{51}$ ergs \cite{supernovae2015}, while the energy of a quark-nova is estimated to be approximately $10^{52}$ ergs \cite{Ouyed_2015, ouyed2022}.
\begin{table}[ht!]
\centering
 \caption{The total and external energy associated with the maximum gravitational mass for both magnetized (with the highest magnetic field) and non-magnetized configurations in each rotating model.}\label{table:2}
\resizebox{0.6\linewidth}{!}{%
 \begin{tabular}{|c|c|c|c|c|}
  \hline
  $\emph{f}$ (Hz)&$B_c$($10^{17}$) G&$M_g (M_\odot)$&$E_{ext}$ ($10^{46}$J)&$E_{tot}$ ($10^{46}$J)\\
  \hline\hline
  \multirow{2}{8pt}{\centering 0}&0&2.35&0.0&54\\
  &5& 2.37&0.05&55\\
  \hline
  \multirow{2}{8pt}{\centering 400}&0&2.38&0.0&54\\
  &5& 2.40&0.05&55\\
  \hline
  \multirow{2}{8pt}{\centering 800}&0&2.48&0.0&54\\
  &5&2.51&0.02&55\\
  \hline
    \multirow{2}{13pt}{\centering 1200}&0&2.73&0.0&55\\
  &5&2.80&0.05&56\\
  \hline
 \end{tabular}}

\end{table}
\subsection{Binding energy and compactness}\label{bind}
\begin{figure}
\centering
\includegraphics[width=0.8\textwidth]{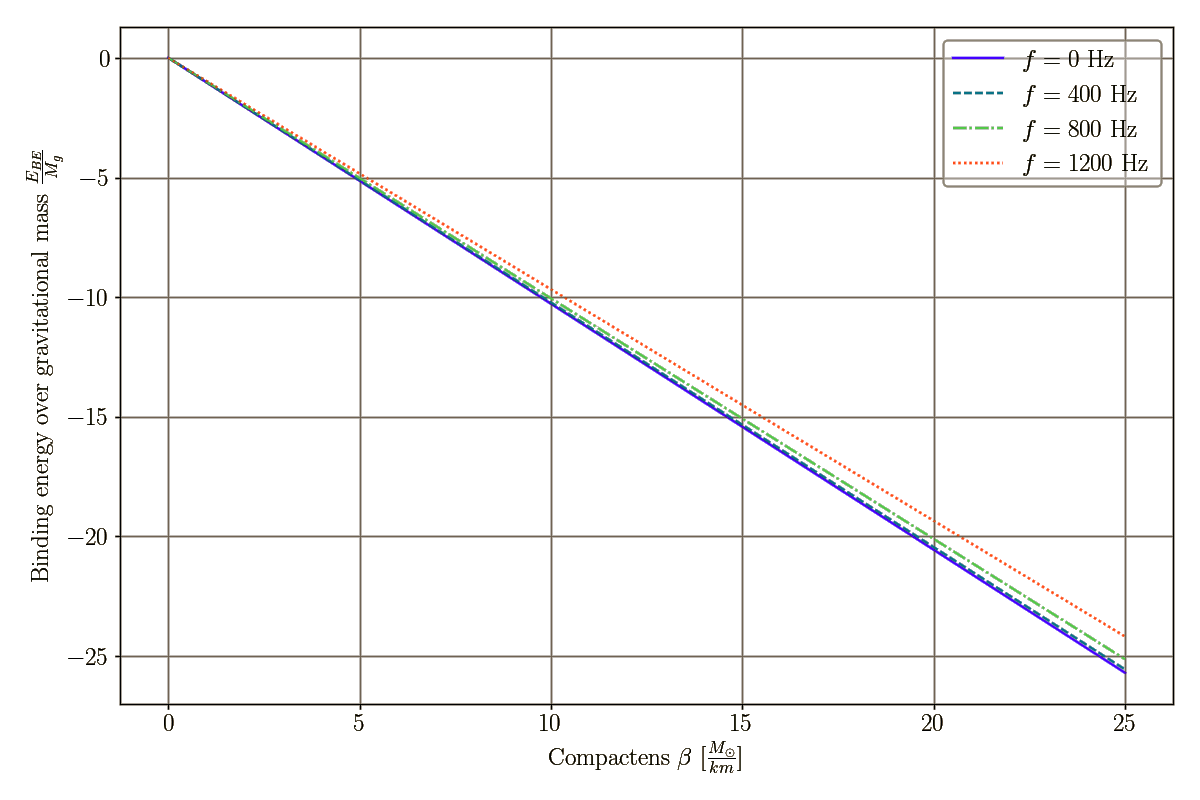}
\caption{Total binding energy per unit of gravitational mass versus compactness.}\label{Compactness}
\end{figure}
An effective and practical way of understanding the microscopic properties of compact objects involves the inverse study of the EOS using observational data. It is possible through the investigation of the relationship between the observable and the theoretical parameters. In this section, we study the relationship between the total binding energy $E_{BE}=(M_g-M_b)c^2+E_{ext}$, where $M_b$ denotes the baryon mass of the star, and the compactness parameter $\beta=M_b/R_{\mathrm{circ}}$ (in units of $M_\odot$/km) in the presented models.

We determined the optimal fit of the binding energy plotted versus the compactness parameter for each rotational frequency and magnetic field. Our analysis indicates that the fitted curves are not significantly influenced by the magnetic field and rotational frequency, as illustrated in Fig. \ref{Compactness}.
Our results show that there is a linear relation between total binding energy and compactness of SQS with the given EOS model:
\begin{equation}
    \frac{E_{BE}}{M_g} = - \beta 
\end{equation}
The binding energy value obtained in our study is slightly smaller than those reported in Refs. \cite{SQSbinding, Drago_2020, Jiang_2019}. For instance, the total binding energy of SQS in a confined density-dependent mass model (CDDM) is $\sim0.41-0.57$, as shown in Table II of \cite{Jiang_2019}. In contrast, the maximum value of $|E_{EB}|/M_g$ in our models is $\sim0.24$. Notably, our value is consistent with the observational data for \emph{J0737-3039B}, \emph{J1756-2251c}, and \emph{J1829+2456c} \cite{Holgado2021}.

In the last column of Table \ref{table:1}, we show that the total binding energy per baryon number is $171 \leq|E_{BE}|/A \leq 184$ Mev. These values of binding energy validate the stability of our models for the compact object.
\section{Conclusions}\label{CONCL}
We investigated the impact of magnetic fields and spin on the structural parameters, stability, and energy of SQS. We constructed a model of a non-magnetized non-rotating SQS, as well as several models of magnetized spinning SQS, by varying the central magnetic field and rotational frequency. In each model, we made a sequence of $51$ configurations by changing the central enthalpy in the range of $0.01 c^2$ to $0.51 c^2$ with a spacing of $0.01 c^2$.

The equation of state of SQS is computed using the density-dependent MIT bag model which takes into account the Landau quantization effect in Fermi relations arising from the strong magnetic field interior of compact objects. To calculate the structural parameters, we used the $3+1$ formalism to solve the axisymmetric Einstein field equations, which resulted in four elliptic partial differential equations.

The LORENE library Et\_magnetisation class was employed to solve the structure equations. We investigated the stability of configurations and examined how the presence of magnetic fields and spin influenced various parameters such as the maximum and minimum gravitational mass, deformation, total energy, binding energy, and compactness.

Our study indicates that the gravitational mass of the SQS is $\approx 2.35 M_\odot$ in a non-rotating model and it slightly increases with the strength of the magnetic field. Furthermore, our analysis of rotating models shows that the star is stable with a larger gravitational mass. Specifically, our model shows a maximum gravitational mass of $2.8 M_\odot$ for a rotation frequency of $\emph{f}=1200$ Hz and a central magnetic field of $B_c\simeq 5\times 10^{17}$ G. We derived a fitting function that relates the maximum gravitational mass to the central magnetic field and rotational frequency (as presented in Section \S\ref{MgR}). 

We also found that in the fast-rotating model, there is a minimum limit for the  gravitational mass (equatorial mass-shedding limit) which is affected by the magnetic field (shown in the last panel of Fig. \ref{MR}). We found that the rate of change in the minimum mass limit is proportional to $\sqrt[4]{B_c}$.  

It is important to note that the results of the minimum mass limit are affected by the computational challenges in the critical model. Therefore, further study is required to confirm the accuracy of the minimum mass value and verify the fitting function.

In addition, we studied the magnetic and rotational deformations of SQS. The deformation parameter is defined as a ratio of the equatorial radius to the polar radius. The results indicate that the maximum deformation of SQS is $1.55$ in the fast-rotating magnetized model with $\emph{f}=1200$ Hz and $B_c= 5\times 10^{17}$ G. We found that the deformation parameter is a function of magnetic field and rotational frequency (discussed in Section \S\ref{Sdeformation}). 

We estimated the total energy of SQS. The total energy of our SQS models is on the order of $10^{54}$ ergs, a finding that is consistent with prior theoretical investigations and observational data. 

Our analysis shows that the binding energy of SQS is a linear function of compactness. In our proposed models, the value of the ratio of binding energy and gravitational mass is approximately $0.24$, which is consistent with observed values for \emph{J0737-3039B}, \emph{J1756-2251c}, and \emph{J1829+2456c}. Moreover, the binding energy per baryon number of SQS is in the range of $171-184$ MeV. The compactness of SQS is approximately $0.25$ across all configurations examined in our study.
\section*{Acknowledgments}
This project was funded by the Polish NCN (grant No. 2019/33/B/ST9/01564), and M\v{C}'s work in Opava also by the ESF projects No. $\mathrm{CZ.02.2.69/0.0/0.0/18\_054/0014696}$. We thank Prof. Włodek Klu\'{z}niak and Prof. Leszek Zdunik for advice and discussions, and the {\sc LORENE} team for the possibility to use the code.
\bibliographystyle{unsrtnat}
\bibliography{references}

\end{document}